\documentclass[aps,twocolumn,showpacs,superscriptaddress]{revtex4}

\usepackage{graphicx}
\usepackage{epsfig}
\usepackage{amssymb}
\usepackage{amsmath}
\usepackage{psfrag}
\usepackage{color}

\newcommand{\br}{{\bf r}}

\newcommand{\dd}{{\mathrm{d}}}

\newcommand{\re}{\mathrm{Re} \, }

\newcommand{\nm}{\,\mathrm{nm}}

\newcommand{\Eu}{\textrm{Eu}^{3+}}

\begin{document}

\title{Mapping and Quantifying Electric and Magnetic Dipole Luminescence at the nanoscale}

\author{L. Aigouy}
\email{lionel.aigouy@espci.fr}
\affiliation{Laboratoire de Physique et d'Etude des Mat\'eriaux, ESPCI Paristech, CNRS, Paris, France}
\author{A. Caz\'e}
\email{acaze@umich.edu}
\affiliation{Department of Mathematics, University of Michigan, Ann Arbor MI 48109, USA}
\affiliation{Institut Langevin, ESPCI ParisTech, CNRS, Paris, France}
\author{P. Gredin}
\affiliation{Institut de Recherche de Chimie Paris, Chimie Paristech, CNRS, Paris, France}
\affiliation{Universit\'e Pierre et Marie Curie, Paris, France}
\author{M. Mortier}
\affiliation{Institut de Recherche de Chimie Paris, Chimie Paristech, CNRS, Paris, France}
\author{R. Carminati}
\email{remi.carminati@espci.fr}
\affiliation{Institut Langevin, ESPCI ParisTech, CNRS, Paris, France}

\begin{abstract}
We report on an experimental technique to quantify the relative importance of electric and magnetic dipole luminescence from a single nanosource in structured
environments. By attaching a $\Eu$-doped nanocrystal to a near-field scanning optical microscope tip, we map the branching ratios associated to two electric dipole
and one magnetic dipole transitions in three dimensions on a gold stripe. 
The relative weights of the electric and magnetic radiative local density of states can be recovered quantitatively, based on a multilevel model.
This paves the way towards the full electric and magnetic characterization of nanostructures for the control of single emitter luminescence.
\end{abstract}


\maketitle


As initially demonstrated by E. Purcell, the luminescence of single emitters is substantially influenced by their local environment~\cite{Purcell46}.
Well-designed nanostructures can use the Purcell effect to enhance or inhibit the spontaneous emission of single nanosources~\cite{Lodahl04, Kuhn06, Anger06}.
In order to achieve an efficient design, a full characterization of the intrinsic photonic properties of nanostructures is required.
The Local Density Of States (LDOS) drives the interaction between a dipole emitter and its environment, and characterizes the dynamics of light
emission or absorption indepedently on the source.
Since the pioneering work of Michaelis {\it et al.}~\cite{Michaelis00}, tremendous progress has been made to use single light-sources attached to the tip of a Near-Field 
Scanning Optical Microscope (NSOM) to probe the LDOS~\cite{Frimmer11, Krachmalnicoff13, Beams13,Cao14}.
All these works have used electric dipole (ED) transitions, {\it e.g.}, in molecules or quantum dots, and therefore only probe the electric part of the LDOS,
the full LDOS containing an additionnal magnetic contribution that can be probed by magnetic dipole transitions~\cite{Joulain03}. 
The control of the electric and magnetic response of nanostructured materials in the optical spectral range has been stimulated by the development of
metamaterials~\cite{Soukoulis07}, or the need for a full characterization of optical antennas~\cite{Kwadrin13}.
Measurements of the magnetic field intensity using engineered tips have been reported~\cite{MagneticMapping}.
The total LDOS has been measured in the infrared domain using Thermal Radiation Scanning Tunnel Microscopy (TRSTM)~\cite{DeWilde06}.
However, due to the difficulty of isolating a single emitter with a dominant magnetic dipole (MD) transition, no direct measurement of the magnetic contribution to the LDOS 
with a resolution on the nanoscale has been achieved so far.

In many situations, light-matter interaction is fully understood by considering electric dipoles, that prevail before higher-order transitions such as magnetic dipoles or electric quadrupoles.
Specific nanostructures such as antennas or resonators can be designed to exit this regime, by exhibiting very intense magnetic fields~\cite{Grosjean11, Hein13, Boudarham14}. 
Rare earth doped emitters can emit dominant MD luminescence when located in the vicinity of a simple gold mirror~\cite{Karaveli10,Karaveli11}.
To describe the competition between two transitions in these kind of sources, the branching ratio is introduced as the relative contribution of a single transition to the total 
luminescence~\cite{Taminiau12, Noginova13}.
In the case of ED and MD transitions, the branching ratios are respectively proportional to the electric and magnetic parts of the radiative LDOS.
Therefore, mapping the branching ratios of a crystal scanned in the near field of a nanostructure would provide a way to fully characterize its
electric and magnetic properties.

In this Letter, we present an experimental technique to map with subwavelength lateral resolution the branching ratios of a $\Eu$-doped nanocrystal in the near field of an arbitrary nanostructure.
We use an NSOM tip to reversibly scan in three-dimensions of space such a crystal and measure the branching ratios associated to one MD and two ED transitions of the emitter versus the distance to a 200 nm thick gold mirror.
We observe similar trends as in Ref.~\cite{Karaveli11}, in great agreement with analytical formulas, thus demonstrating the robustness of the method.
We present three-dimensional maps of the branching ratios in the vicinity of a 2 $\mu$m wide gold stripe, and identify areas where the ED and MD transitions successively dominate the luminescence.
The knowledge of these maps permits the tuning of the luminescence of ED and MD transitions using appropriate positioning on the nanostructure.
Finally, we recover from experimental data the relative electric and magnetic parts of the radiative LDOS using a multilevel model, in quantitative agreement with theory.
This technique will be of great interest in the growing field of engineering nanostructures for the control of spontaneous emission.

\begin{figure}
\begin{center}
\includegraphics[width=0.4\textwidth]{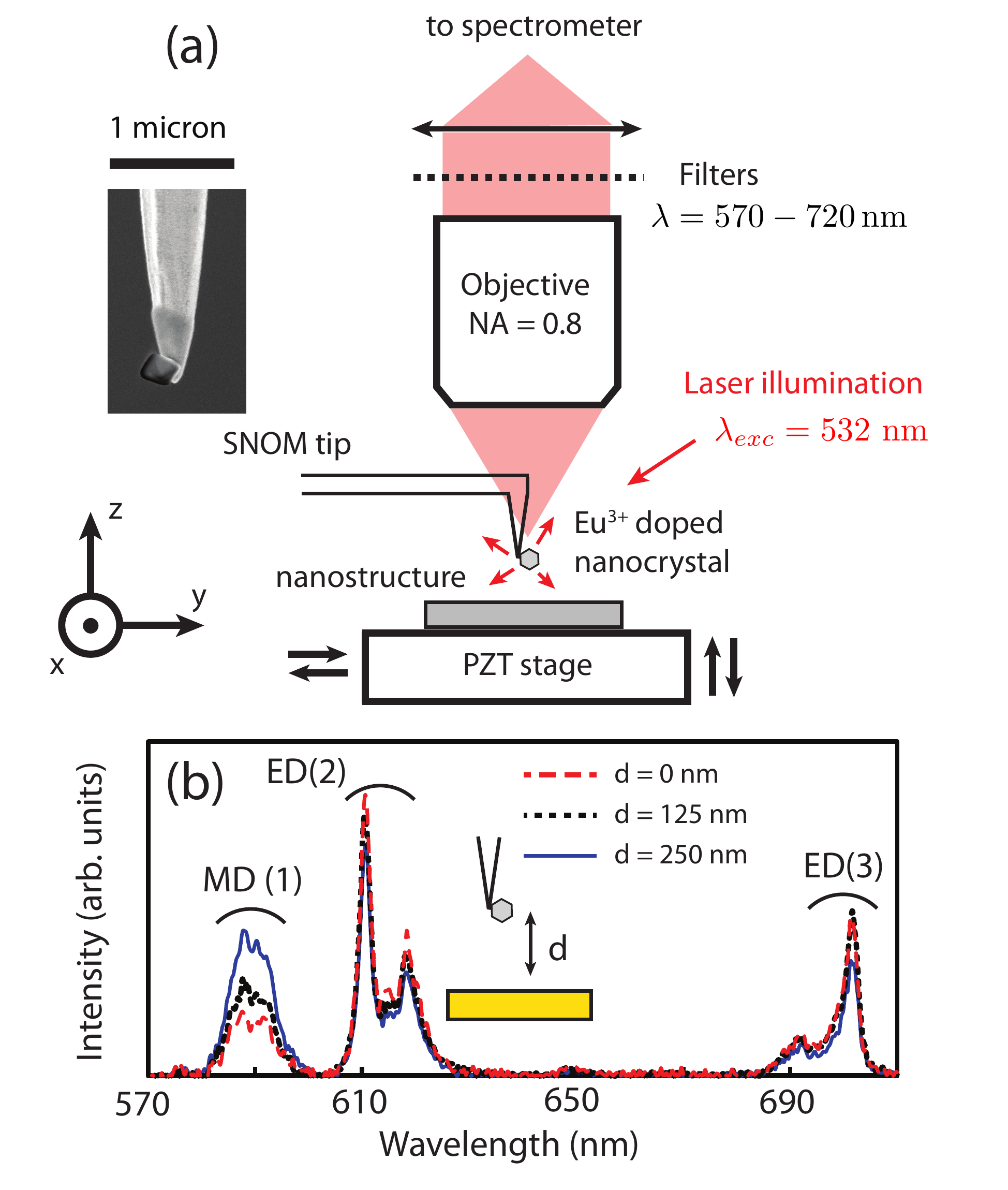}
\caption{\label{fig:fig1} (a) Experimental setup. (Inset) Scanning Electron Microscope image of the $\Eu$-doped nanocrystal attached to the tungstene NSOM tip. The nanocrystal is approximately $200\,\nm$ large. (b) Luminescence spectra of the $\Eu$-doped KYF nanocrystal at several distances from a gold mirror.}
\end{center}
\vspace{-0.5cm}
\end{figure}

\begin{figure}
\begin{center}
\includegraphics[width=0.35\textwidth]{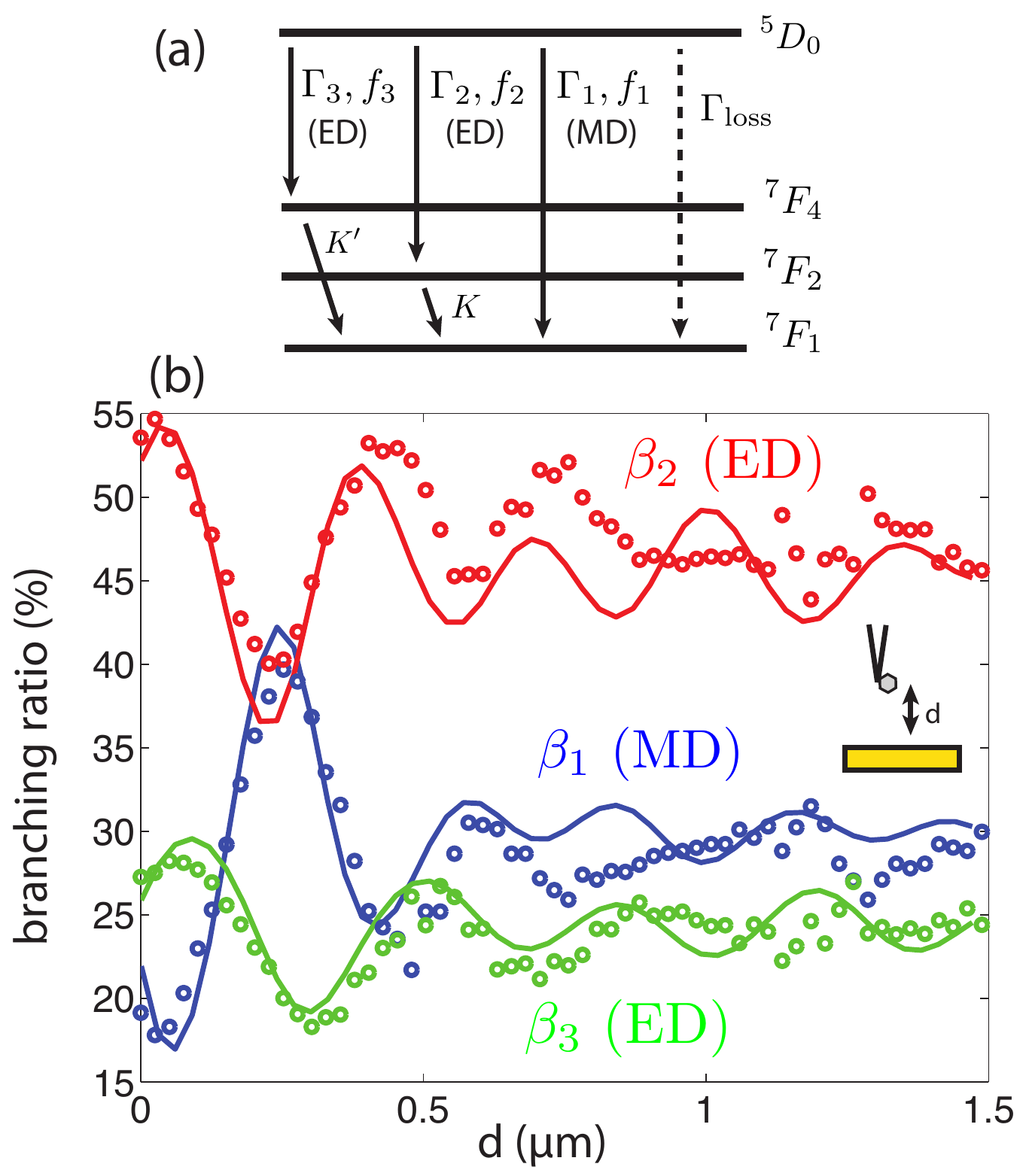}
\caption{\label{fig:fig2} (a) Band diagram of the $\Eu$ ions. (b) Branching ratios associated to transitions 1 (blue), 2 (red) and 3 (green) as a function of the distance to a gold mirror. (Full lines) Analytical expressions; (Circles) Experimental values.}
\end{center}
\vspace{-0.5cm}
\end{figure}

%
The experimental set-up is depicted in Fig.~\ref{fig:fig1}(a). 
%
KYF nanocrystals, doped with $\Eu$ ions, were synthesized by coprecipitation of an aqueous solution of potassium and yttrium nitrates in hydrofluoric acid at R.T.
(where KYF is KY7F22 crystal phase).
Details of the synthesis will be published elsewhere. 
A single crystal was glued at the end of a sharp tungsten tip with a nanomanipulation system. 
It was mounted on a homemade atomic force microscope combined with a classical optical setup for ensuring luminescence excitation and detection. 
Illumination is made at oblique incidence ($\lambda = 532\,\mathrm{nm}$) and luminescence is collected with a high numerical aperture objective (NA = 0.8), situated above the tip and the sample~\cite{Wang09,Aigouy11}. 
Light is then sent to a spectrometer coupled with a cooled CCD camera. 
For this configuration, the sample can move in all three directions of space and the tip is immobile. 
The tapping mode (with oscillation amplitude $\sim 20$ nm) was used to control the tip-sample distance. 
Typical emission spectra of the nanocrystal at three distances from a 200 nm thick gold mirror are shown in Fig.~\ref{fig:fig1}(b). 
These spectra exhibit three dominant peaks, which intensities exhibit different variations with respect to the distance to the mirror.
Those peaks can be associated to radiative transitions in a four-level model of the $\Eu$ ion, as shown in Fig.~\ref{fig:fig2}(a)~\cite{Karaveli10}.
The peak located between 580 nm and 600 nm is connected to the MD transition $^5D_0$ $\rightarrow$ $^7F_1$ and will be labelled by 1. 
The peaks in spectral ranges 600-630 nm and 685-705 nm are linked, respectively, to the two ED transitions $^5D_0 \rightarrow ^7F_2$ and $^5D_0$ $\rightarrow$ $^7F_4$ and are labeled by 2 and 3.
Note that the transition $^5D_0 \rightarrow ^7F_3$ (peak 645-655 nm) is very weak, and is ignored in the analysis.
%
%
%
%
To quantify the relative contribution of transition $j$ to the total luminescence, one can introduce the branching ratio $\beta_j(\br)$, defined as~\cite{Karaveli10}
\begin{equation}
\beta_j(\br) = I^{fluo}_j(\br)/I^{fluo}_{total}(\br).
\end{equation}
where $\br$ is the location of the emitter, $I^{fluo}_j(\br)$ is the fluorescence intensity detected in the spectral range of transition $j$ , and $I^{fluo}_{total}(\br)$ is the total fluorescence intensity.
We have measured the branching ratios of transitions 1, 2 and 3 versus the distance to the gold mirror.
Results are displayed in Fig.~\ref{fig:fig2}(b).
One can observe oscillations, that are maximum in amplitude at short distance from the surface.
At long distance, all branching ratios tend to stabilize to their value in vacuum. 
We recover the same trends as that reported in Refs.~\cite{Karaveli10,Karaveli11}.
The slight difference is due to the fact that in our setup no spacer is used between the metal surface and the emitter, and the luminescence is emitted
from a single sub-wavelength-sized particle.
The oscillations are due to the interferences between incident and reflected modes on the surface. 
Importantly, those oscillations exhibit different trends for ED and MD transitions.

To understand the experimental results, we use the four-level model represented in Fig.~\ref{fig:fig2}(a).
The radiative transition rates are denoted by $\Gamma_1$, $\Gamma_2$ and $\Gamma_3$.
Two non-radiative transitions to the most stable state $^7F_1$ are introduced, with rates $K$ and $K'$.
These transitions are associated with internal vibrational modes, and are considered instantaneous compared to all radiative decays.
A phenomenological transition rate $\Gamma_{loss}$ is added to take into account non-detected radiative and other non-radiative decays.
The rate equations in steady-state yield $N(^7F_4) = N(^7F_2) = 0$, and
\begin{equation}
N(^5D_0) = \frac{A\sigma_{abs}I_{inc}}{\Gamma_1+\Gamma_2+\Gamma_3 + \Gamma_{loss}}\,N(^7F_1),
\end{equation}
where $N(S)$ is the population of state $S$, $\sigma_{abs}$ the source absorption cross section, $I_{inc}$ is the local intensity at its position and $A$ a proportionality constant characteristic of the experimental setup.
The fluorescence signal associated to each radiative transition is proportional to the corresponding decay rate $I^{fluo}_j(\br) = N(^5D_0) \,\Gamma_j$.
Importantly, the proportionality constant is independent on the transition.
Assuming that most of the detected fluorescence comes from transitions 1, 2 and 3, the branching ratio can be approximated by
\begin{equation}
\beta_j(\br) = \frac{I^{fluo}_j(\br)}{\sum_j I^{fluo}_j(\br)} = \frac{\Gamma_j}{\Gamma_1 + \Gamma_2 + \Gamma_3}.
\end{equation}
In these conditions, the branching ratio is independent on the illumination. 
We have verified this experimentally by changing the polarization of the incident light without varying the curves shown in Fig.~\ref{fig:fig2}(b).
Note that the branching ratios do not depend on the phenomenological transition rate $\Gamma_{loss}$. 
We assume all levels of the emitter to be spectrally narrow, so that the radiative transition rates are driven by the electric or magnetic radiative LDOS at one frequency $\omega_j = 2\pi c/\lambda_j$, $\lambda_j$ being the central wavelength associated to transition $j$.
In all calculations, we have assumed the wavelength of the radiative transitions to be $\lambda_1 = 590 \,\mathrm{nm}$, $\lambda_2 = 615 \,\mathrm{nm}$ and $\lambda_3 = 695 \,\mathrm{nm}$.
Decay rates of radiative ED (resp. MD) transitions are proportional to the radiative electric (resp. magnetic) LDOS $\rho_E^R(\br,\omega)$ (resp. $\rho_M^R(\br,\omega)$). 
In order to take into account intrinsic properties of the crystal, we introduce oscillator strengths for each transitions $f_j$, such that $\Gamma_j \propto f_j \rho^R_{E/M}(\br,\omega_j)$.
The branching ratio is connected to the electric (magnetic) radiative LDOS via
\begin{equation}
\label{jeanlouis}
\beta_j = \frac{f_j\rho_{E/M}^R(\br,\omega_j)}{f_1\rho^R_M(\br,\omega_1)+f_2\rho^R_E(\br,\omega_2)+f_3\rho^R_E(\br,\omega_3)}.
\end{equation}
Far enough from any structure, the electric and magnetic part of the radiative LDOS become equal~\cite{Joulain03}.
Assuming that $f_1+ f_2 + f_3 = 1$, we have used the value of the branching ratio $\beta_j$ measured at large distance to the mirror as an estimate of $f_j$.
The gold mirror is modeled by a semi-infinite half space $z<0$ filled with gold (dielectric constant taken from Ref.~\cite{PalikBook}), separated from a semi-infinite
vacuum half space $z>0$.
The radiative part of the electric LDOS at frequency $\omega$ and distance $z>0$ from the metal reads~\cite{Joulain03}
\begin{equation}
\label{electric_ldos}
\begin{split}
\frac{\rho^R_E(z,\omega)}{\rho_v}= \int_{0}^{\mathrm{NA}}\frac{\kappa\,\dd\kappa}{4p} \left[ 2 + \re \left(r_{12}^s \exp(2ip\,\omega z/c)\right) \right. \\
\left. + \re \left(r_{12}^p \exp(2ip\,\omega z/c)\right) \left(2\kappa^2-1\right) \right],
\end{split}
\end{equation}
where $\rho_v$ is the total LDOS in vacuum, $p=\sqrt{1-\kappa^2}$ and
$r_{12}^{s,p}$ are the Fresnel reflexion coefficients at the gold/air interface for $s$ and $p$ polarizations,
respectively~\cite{Joulain03}.
The radiative part of the magnetic LDOS is obtained by switching $r_{12}^s$ and $r_{12}^p$ in Eq.~(\ref{electric_ldos}).
The upper bound in the integral in Eq.~(\ref{electric_ldos}) is the numerical aperture NA = 0.8 defined by the detection setup.
The theoretical calculations are compared to experiments in Fig.~\ref{fig:fig2}(b). 
To account for the finite size of the nanocrystal, the branching ratios were averaged over emitter position in a cubic volume with size 180 nm. 
The calculations are in excellent agreement with the experimental data. 
In particular, the period and amplitude of the oscillations are recovered. 
Note that apart from the size of the crystal, no adjustable parameter is included in the model.

\begin{figure}
\begin{center}
\includegraphics[width=0.5\textwidth]{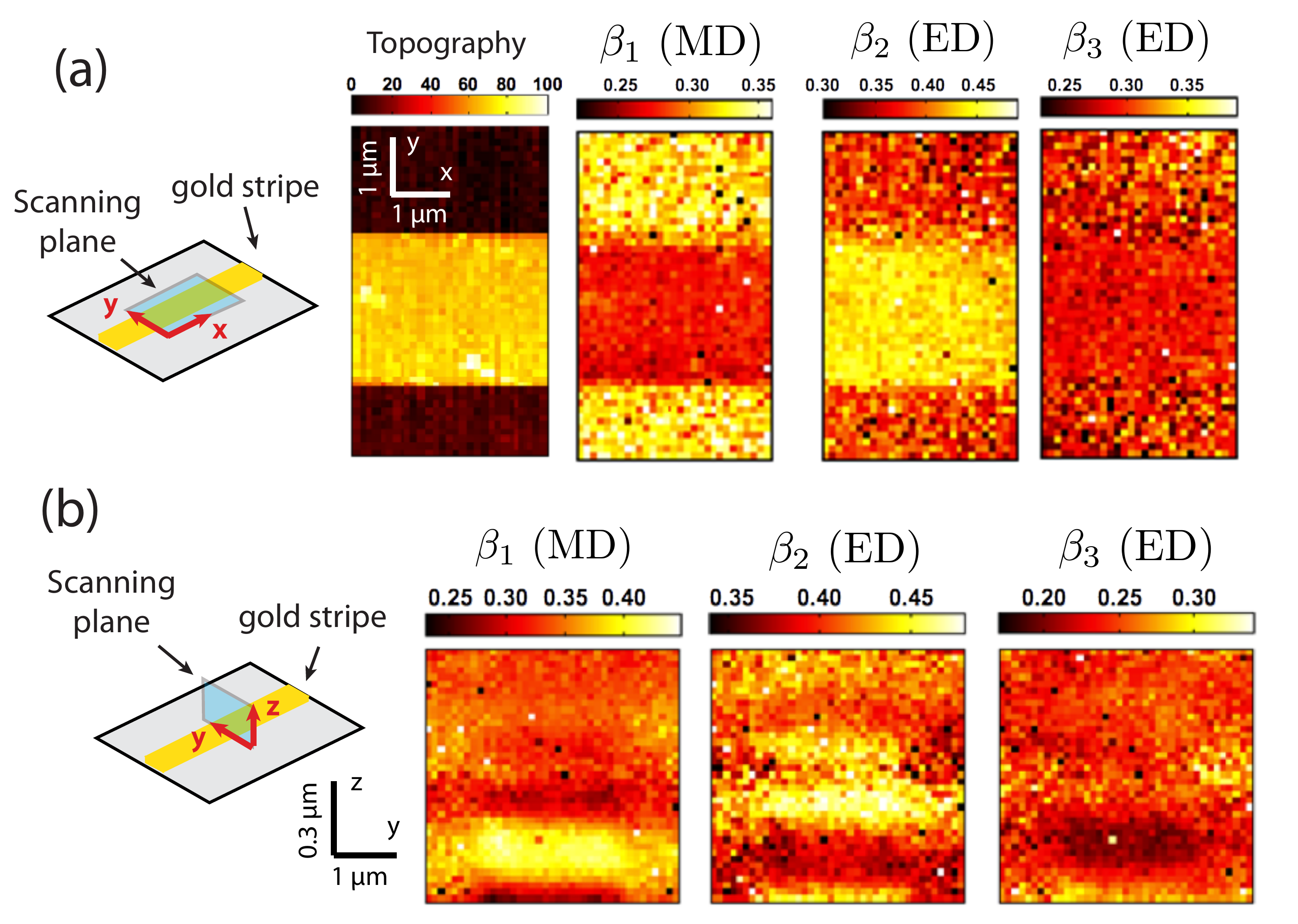}
\caption{\label{fig:fig3} : (a) (From left to right) Topography of the gold stripe; Branching ratios of transitions 1, 2 and 3 in the $x-y$ plane. Image size is $3 \times 4.8 \,\mu m^2$. (b) (From left to right) Branching ratios of transitions 1, 2 and 3 in the $y-z$ plane. Image size is $4\times 1\,\mu m^2$.}
\end{center}
\vspace{-0.5cm}
\end{figure}

%
The use of an NSOM in our experimental setup allows to measure three-dimensional maps of the branching ratios.
To demonstrate this imaging capability, we have chosen a structure made of a gold stripe fabricated on a SiO$_2$ substrate. 
The stripe is 2 $\mu$m-wide, 30 $\mu$m-long and 60 nm-thick. 
We first performed a scan of the stripe in contact mode. 
The topography measured by the AFM and the branching ratios maps measured in the $x-y$ plane, parallel to the stripe, are shown in Fig.~\ref{fig:fig3}(a). 
%
We observe that the branching ratio of the MD transition is larger on SiO$_2$ than on the Au stripe. 
Transition 2 (ED) has an opposite behavior, in analogy with what was already observed in Fig.~\ref{fig:fig2}(b).
The weak contrast does not allow one to make any conclusion concerning transition 3 (ED), although the trends should be analog to those of transition 2. 
These maps can be explained by the optical properties of Au and SiO$_2$, which have different reflection coefficients. 
As in the case of the approach curves in Fig.~\ref{fig:fig2}(b), the contrast is driven by the interference between incident and reflected modes.
Since Au is highly reflective and SiO$_2$ is transparent, the interference is different on the two materials, giving rise to the observed contrast~\cite{note_contrast}. 

We show in Fig.~\ref{fig:fig3}(b) the branching ratios measured when scanning above the stripe, in the $y-z$ plane, perpendicular to the surface. 
First, the most intense variations of the branching ratios occur above gold. 
In this region, the trends are analog to those observed when approaching the gold mirror in Fig.~\ref{fig:fig2}(b).
The ED transitions dominate at short distance from the stripe. 
As the distance increases, the ED and MD dominate alternatively the luminescence, as commented above.
The trends on SiO$_2$ are similar, but weaker, due to the smaller value of the reflection coefficient on silica.
These maps demonstrate that the near field of an inhomogeneous object exhibits rich variations of the branching ratios in all three directions of space.
These experiments illustrate the necessity for a complete near-field characterization of nanostructures.

\begin{figure}
\begin{center}
\includegraphics[width=0.4\textwidth]{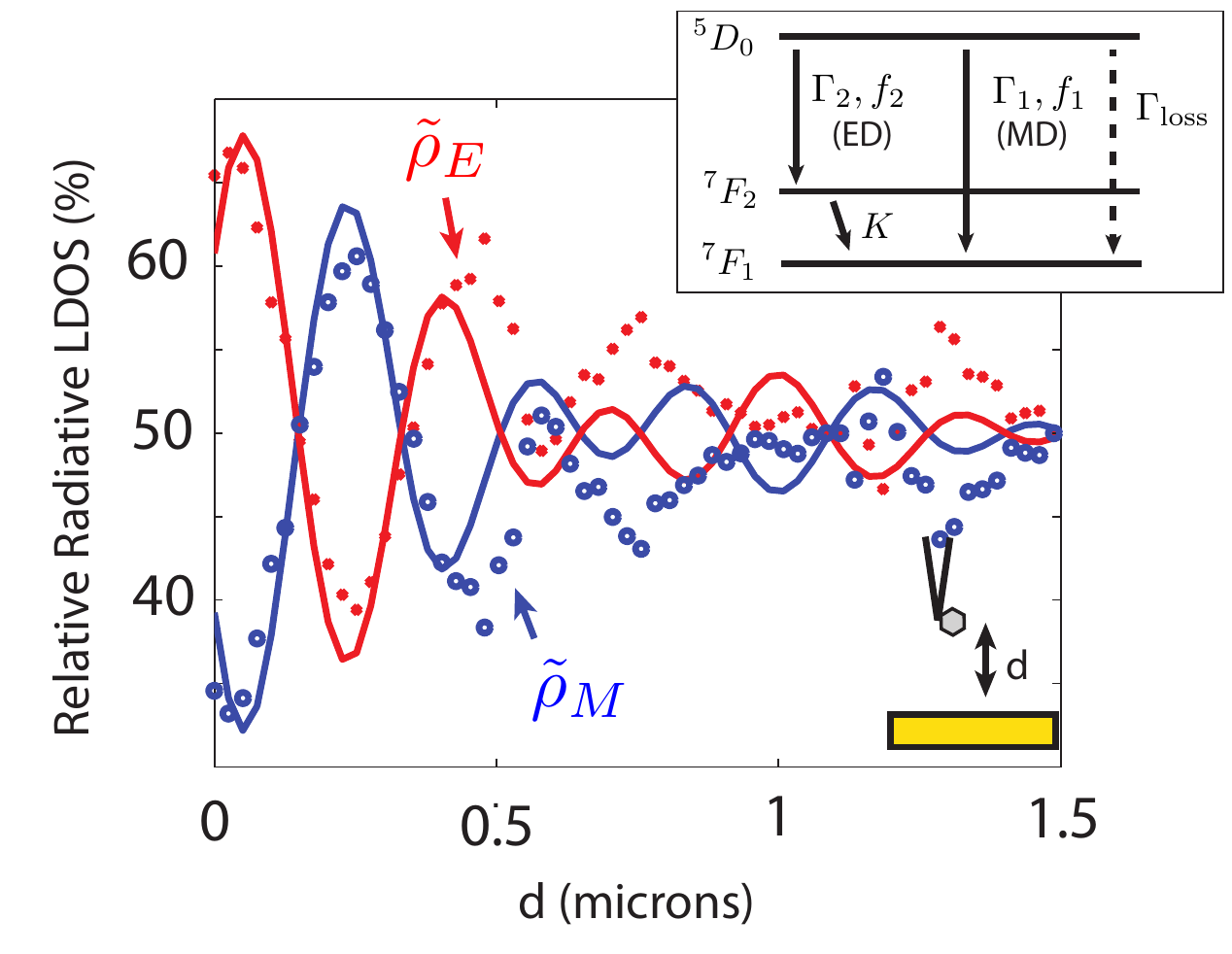}
\caption{\label{fig:fig4} (a) Three-level model of $\Eu$ ions; (b) Relative radiative LDOS versus the distance to a gold mirror; (Full line) Analytical formulas; (Red dots) Experimental relative electric radiative LDOS; (Blue circles) Experimental relative magnetic radiative LDOS.}
\end{center}
\vspace{-0.5cm}
\end{figure}

%
Last but not least, we have used the method first proposed in Ref.~\cite{Taminiau12} to recover the relative electric radiative LDOS, defined as
\begin{equation}
\label{jeanmichel}
\tilde{\rho}_E(\br) = \frac{\rho_E(\br,\omega_2)}{\rho_E(\br,\omega_2) + \rho_M(\br,\omega_1)},
\end{equation}
where $\rho_E(\br,\omega_2)$ is the electric LDOS at frequency $\omega_2$, and $\rho_M(\br,\omega_1)$ the magnetic LDOS at frequency $\omega_1$, both at position $\br$ of the crystal.
The analog quantity $\tilde{\rho}_M = 1 - \tilde{\rho}_E$ is called the relative magnetic radiative LDOS.
While the branching ratio is the relevant quantity to measure the relative importance of two competing transitions for one particular emitter, the two relative radiative LDOS quantify the competition between ED and MD luminescence independently on the nature of the source. 
In terms of the present model, $\tilde{\rho}_E$ and $\tilde{\rho}_M$ are independent on the oscillator strengths of the transitions.
One can see the distinction by comparing Eqs.~(\ref{jeanlouis}) and~(\ref{jeanmichel}).
To directly measure $\tilde{\rho}_E$ and $\tilde{\rho}_M$ , one needs an emitter with an ED and a MD transition that share a similar oscillator strength.
Here, the multilevel model allows to use any source with an ED and a MD transition.
Let us consider the simplified model of $\Eu$ luminescence sketched in Fig.~\ref{fig:fig4}(a). 
Luminescence from transition 3 is assumed non-detected and is incorporated in the phenomenological rate $\Gamma_{loss}$.
We define the new branching ratios $\beta'_1 = \Gamma_1 / (\Gamma_1 + \Gamma_2)$ and $\beta'_2 = \Gamma_2 / (\Gamma_1 + \Gamma_2)$, that can be deduced straightforwardly from the experimental data. 
The relative electric radiative LDOS is connected to the new branching ratios via
\begin{equation}
\tilde{\rho}_E(\br) = \frac{\beta'_2f_1}{\beta'_2f_1+\beta'_1f_2}
\end{equation}
The relative magnetic radiative LDOS is directly deduced from $\tilde{\rho}_M = 1 - \tilde{\rho}_E$.
Using the oscillator strength obtained earlier, we recover the relative electric and magnetic radiative LDOS from the experimental data plotted in Fig.~\ref{fig:fig2}(b).
We compare those values to their theoretical expressions in Fig.~\ref{fig:fig4}.
%
The agreement between theory and experiment is excellent at short distances from the mirror.
At long distances, the oscillations of the branching ratios become smaller, reducing the efficiency of the recovery.
However, Fig.~\ref{fig:fig4} demonstrates unambiguously the relevance of this technique to measure the relative electric and magnetic radiative LDOS.
Access to those quantities could be of great interest in the growing field of engineering nanostructures for the control of single emitters luminescence.

%
In summary, using a single $\Eu$-doped nanocrystal glued at the end of a sharp tip, we have developed a scanning probe that allows to map simultaneously the branching ratios associated to a MD and two ED transitions in three directions of space in the near field of nanostructures. 
We have demonstrated the robustness of this technique by comparing it with analytical formulas, with no adjustable parameter except the size of the crystal.
We have presented three-dimensional maps in the near field of a 2 $\mu$m wide gold stripe, exhibiting areas where the ED and MD transitions successively dominate the far-field luminescence.
Using a multi-level model, we have shown that the relative electric and magnetic parts of the radiative LDOS could be recovered from the experimental data.
This technique should reveal very useful to reach a full characterization of the near-field properties of structures like nano-antenna or split-ring resonators, that exhibit strong magnetic fields. 
This work paves the way towards new ways of engineering nanostructures for the control of the luminescence of single emitters.

This work is supported by LABEX WIFI  (Laboratory of Excellence within the French Program "Investments for the Future") under references ANR-10-LABX-24 and ANR-10-IDEX-0001-02 PSL*, by Dim NanoK through projects NanoFEG (2012) and Snalbun (2008), and by CNRS through the project ``Instrumentation aux limites''.



\begin{thebibliography}{40}
\bibitem{note_contrast}
This contrast is observed with an approximately $200\,\mathrm{nm}$-diameter particle, so all the signals are averaged over this volume. A different contrast could be observed near the surface with a smaller nanocrystal.
\end{thebibliography}
\end{document}